# Influence of defects on the lattice constant of GaMnAs

J. Sadowski [1,2], J. Z. Domagala [2]

[1] *MAX-Lab, Lund University, Po. Box. 118, SE-221 00 Lund, Sweden,*

[2] *Institute of Physics, Polish Academy of Sciences, Al. Lotników 32/46, PL-02-668 Warszawa, Poland*

We study the influence of main compensating defects: As antisites and Mn interstitials, known to occur in GaMnAs ferromagnetic semiconductor, on its structural properties. Our experimental results show that there is a balance between Mn interstitial and As antisite defects, leading to the reduced density of one type of defect upon increased density of another defect. The significant differences in the lattice parameters of GaMnAs with the different balance between these two types of defects were observed. The annealing induced reduction of GaMnAs lattice constant is inhibited in the samples with large density of As antisites.

Ferromagnetic semiconductors (FMS) witnessed a considerable increase of the research activity in recent years. Although known for quite a long time - for example in EuS, EuO[1], IV-VI narrow gap materials alloyed with Mn (PbSnMnTe)[2]; ferromagnetism in semiconductors has gained a renewed interest due to the prospects of using these materials in the new kind of magnetoelectronic (spintronic) devices. The advent of III-V FMS: InMnAs[3] and GaMnAs[4] in 1992 and 1996 respectively, generated a new momentum in the research activity in this area, due to their compatibility with the existing III-V semiconductor technology. The extensive research activity in the field caused the considerable progress in both: understanding the physical phenomena leading to the ferromagnetism in III-V FMS[5–10] and improving the magnetotransport properties of these materials[11–15]. This, very recently led to the increase of Tc from the previously established limit of 110 K[2] to 160 – 170 K[11,12,16]. Different theoretical approaches concerning FMS foresee considerably higher Tc[5–10], so further progress in that direction can not be excluded. The recent advancements in increasing Tc in GaMnAs were possible due to recognizing the most important defects compensating Mn acceptors. However, in contrast to the previous studies[17–19], where only As antisites ($As_{Ga}$) were considered, nowadays only Mn interstitial defects ($Mn_I$), recently verified experimentally to occur in GaMnAs[20], are taken into account[21]. On the other hand, it is obvious that both $Mn_I$ and $As_{Ga}$ are present in GaMnAs. Moreover concentrations of these two defects are expected to be close to each other, 0.1% - 1% (or even 1.75% as suggested by some theoretical works[19]) for $As_{Ga}$ and up to 2% for $Mn_I$[22]. As shown by several groups[20,23-26] the control over $Mn_I$ defects is possible via the post-growth annealing procedures. It was demonstrated[20] that the post growth annealing reduces the concentration of $Mn_I$ defects in the volume of GaMnAs layers, though it is not clear what happens to the Mn atoms removed from the interstitial positions. There are some suggestions that they segregate at the GaMnAs surface[27,28].

As reported by Yu et al.[20] it is possible to detect $Mn_I$ atoms directly, by the particle induce X-ray emission and Rutheford back-scattering methods. As concerns As antisites in GaMnAs – no experimental data revealing their concentrations are available yet. On the other hand, the As antisite defects in GaAs grown by low temperature Molecular Beam Epitaxy (MBE) have already been investigated in details[29-31] and it is well known how to estimate their content. Typical methods used for evaluation of $As_{Ga}$ in LT GaAs, namely lattice constant measurements and optical absorption/emission measurements are difficult in the case of GaMnAs. The use of optical methods is complicated due to the very poor optical quality of this compound. The straightforward information on $As_{Ga}$ concentration from the lattice constant measurements is hard to obtain since the GaMnAs lattice expansion is due to the several factors, such as: Mn at Ga sites; Mn at interstitial sites; As at Ga sites[32]; other defects typical for LT GaAs. As is well known[29,30], the density of As antisites in LT GaAs ($[As_{Ga}]$) depends on the MBE growth conditions and can be adjusted by varying either the substrate temperature ($T_s$) or the As to Ga flux ratio. Increasing $T_s$ decreases $[As_{Ga}]$, whereas increasing As/Ga increases $[As_{Ga}]$[30]. In this work we used the latter way to change $[As_{Ga}]$ in both LT GaAs buffers and subsequently grown GaMnAs layers. We separate the influence of $Mn_I$ and $As_{Ga}$ defects on the lattice parameter $a_{GaMnAs}$ by careful X-ray diffraction (XRD) measurements of $Ga_{0.96}Mn_{0.04}As$ samples, differing in the concentrations of As antisites.

We have investigated three sets of $Ga_{0.96}Mn_{0.04}As$ layers grown on LT GaAs buffer layers, and a sequence of GaMnAs layers with Mn content increasing from 0.1% to 3.5%. The samples were grown in the KRYOVAK MBE system dedicated to III-Mn-V magnetic semiconductors. A valved cracker source was used to generate $As_2$ flux. Before the growth of $Ga_{0.96}Mn_{0.04}As$ layers the LT GaAs buffers were deposited at the same substrate temperature (230 $^o$C) and growth rate (0.2 ML/s) as further used for GaMnAs. Each of these samples was grown at the different $As_2$ / Ga flux ratio and the other parameters like $T_s$ and Mn content – the same. Samples 1, 2 and 3 were grown at $As_2$ to Ga flux ratios of 2, 5 and 9, respectively. The Mn content was set by the temperature of Mn effusion cell $T_{Mn}$ = 775 $^o$C, the same for all three samples, and verified by measuring the differences in the growth rates between GaMnAs and LT GaAs, using the RHEED intensity oscillations[33]. After the MBE growth the substrate temperature was decreased rapidly, the samples were taken out of the vacuum system and cleaved into 4 pieces. One piece was left unchanged; the other pieces were placed on the molybdenum holder again, put into the vacuum system and transferred to the MBE growth chamber for annealing at high vacuum. The annealings were performed in such a way that at each annealing run the pieces of all three samples were annealed together. Annealings at different temperatures were done for different pieces of each sample. The annealing temperatures were chosen to be 240, 260 and 280 °C, the annealing time was 2 hours in each case.

For XRD measurements, we used a PHILIPS X-pert high-resolution diffractometer with the collimating mirror. Samples were measured in the two different configurations: double axis, for omega and omega/2theta scans; triple axis with an analyzer, for 2 theta/omega scans and reciprocal space mapping. Both symmetrical and asymmetrical Bragg reflections were measured. Our investigations of GaMnAs structure by the XRD methods were inspired by the observation of significant influence of the growth conditions, namely As to Ga flux ratio and substrate temperature on the GaMnAs lattice constant. This is illustrated in Fig. 1. Two GaMnAs samples measured by XRD have the same Mn content of 0.1%, they were grown at the same substrate temperature (230 °C) and differ in the $As_2$ to Ga flux ratio, which is 2 for sample (a) and 9 for sample (b). The clear difference in angular positions of GaMnAs (004) diffraction peaks, reflecting the differences between the lattice parameters of the two samples can be seen. Similar effect occurs in GaMnAs with higher Mn content. Our observations are consistent with the results published by other groups. Shott et al.[34] investigated the effect of the substrate temperature on the GaMnAs lattice parameter. The authors observed significant changes of the GaMnAs lattice constant for the samples with the same Mn content grown at different $T_s$. These observations make questionable the sense of extrapolating the GaMnAs lattice constant to the zinc blende MnAs as well as the estimations of Mn content in GaMnAs from the lattice constant measurements.

Another interesting feature concerning the GaMnAs lattice parameter is illustrated in Fig.2. This figure shows the dependence of strained, perpendicular GaMnAs lattice constant on the Mn content, starting from the diluted samples containing 0.1% Mn, up to the Mn content ($X_{Mn}$) of 3.5%. It is interesting to notice that $a_{GaMnAs}$ *decreases* with increasing $X_{Mn}$ at the very low Mn concentration range. For Mn composition increasing form 0.1% to 0.3%, $a_{GaMnAs}$ decreases, reaches the minimum value at about 0.3% Mn and then increases proportionally to $X_{Mn}$. This decrease of the GaMnAs lattice constant in the much-diluted Mn concentration limit was not reported before, to our best knowledge. It is discussed in the further part of the paper.

Figure 3 shows the XRD results for three $Ga_{0.96}Mn_{0.04}As$ layers grown on thick LT GaAs buffers. The most interesting aspects of Fig. 3 are the dependences of angular positions of (006) Bragg reflections of LT GaAs buffers and GaMnAs layers on $As_2$/Ga flux ratio ($r_{As/Ga}$). These Bragg reflections are measure of a perpendicular, strained lattice parameter of the epilayers. In the sample 1 grown at $r_{As/Ga} = 2$ the relaxed lattice constant of LT GaAs buffer (calculated from measured strained lattice parameter) is the lowest for all three samples: $a_{LTGaAs}(1) = 5.6549$ Å, whereas the lattice constant of GaMnAs – $a_{GaMnAs}$ is the highest. In samples 2 and 3 grown and medium and high $r_{As/Ga}$, respectively, $a_{LTGaAs}$ increases with increasing $r_{As/Ga}$, but $a_{GaMnAs}$ changes slightly in the opposite way than $a_{LTGaAs}$, i.e. it's the largest for sample 1, and smaller for samples 2 and 3.

As reported by many groups, the changes in the LT GaAs lattice constant are caused by the different densities of As antisites. Following the relation between $a_{LTGaAs}$ and density of $As_{Ga}$ defects given by Liu et. al.[29] we estimated [$As_{Ga}$] to be: 0.20%, 0.23% and 0.35% for samples 1, 2 and 3, respectively. Assuming that [$As_{Ga}$] in GaMnAs is the same as in LT GaAs buffers grown prior to the GaMnAs deposition and knowing $a_{GaMnAs}$ from measurements, we can estimate the contribution of Mn interstitial defects to $a_{GaMnAs}$. We are using the results of theoretical approach of Masek et. al.[32], who calculated the dependence of $a_{GaMnAs}$ on [$As_{Ga}$], [$Mn_I$] and Mn at Ga sites to follow the formula:

$$a_{GaMnAs}(x,y,z) = a_0 + 0.02x + 0.69y + 1.05z \quad (1)$$

where: $a_0$ – lattice constant of GaAs without defects; x – concentration of Mn at Ga sites; y – concentration of As antisites; z – concentration of Mn at interstitial sites.

Using the values of [As$_{Ga}$] calculated form the lattice parameters of LT GaAs buffers and the measured values of $a_{GaMnAs}$ for samples 1, 2 and 3, we obtain the following concentrations of Mn$_I$ (z in equation 1):

sample 1: [Mn$_I$] = 0.80% for [As$_{Ga}$] = 0.20%

sample 2: [Mn$_I$] = 0.68% for [As$_{Ga}$] = 0.23%

sample 3: [Mn$_I$] = 0.61% for [As$_{Ga}$] = 0.35%

The numerical parameters obtained from the results of XRD measurements shown in Fig. 3 and used for calculations of [Mn$_I$] from formula (1) are given in Table 1.

Assuming that $a_{GaMnAs}$ dependence on x, y, z is following equation (1), we can explain the decrease of $a_{GaMnAs}$ at low Mn content region, shown in Fig. 1. If Mn atoms are introduced into the LT GaAs lattice in a very small amount (below 0.3% in our case) they enter only the Ga sites. Since the concentration of As$_{Ga}$ is higher than the Mn content, the system does not need any additional compensating defects other than As antisites, to overcompensate Mn acceptors. Increasing Mn content up to the [As$_{Ga}$] value (0.1% - 0.5% depending on the LT MBE growth conditions) finally leads to the situation, when the concentration of Mn acceptors is higher than the concentration of As antisite donors and another kind of defect is necessary to compensate Mn acceptors. These defects may be Mn at interstitial positions. Following formula (1), in the low Mn concentration limit and below the compensation point, we expect [As$_{Ga}$] to be constant, [Mn$_I$] to be equal to zero. This gives the increase of the lattice constant of 0.00002 Å, for GaMnAs containing 0.1% Mn. This is below the resolution of a typical XRD setup. That means that Mn should not influence the measured value of GaMnAs lattice constant in the case when all the Mn atoms are situated at the Ga sites, in this low Mn content range. The slight decrease of $a_{GaMnAs}$ with increasing $X_{Mn}$ at 0.1% < $X_{Mn}$ <

0.3% can be caused by the lattice contraction due to the ionization of deep $As_{Ga}$ donors by Mn acceptors. Similar effects were observed by Specht et. al.[35] in LT GaAs doped to p-type by Be and C. The authors observed the ionization of As antisites proportional to the concentration of p-type dopands and concluded that ionized $As_{Ga}$ defects have different (smaller) sizes than neutral $As_{Ga}$. Attributing the $a_{GaMnAs}$ ($X_{Mn}$) lattice constant minimum to the 100% ionization of $As_{Ga}$ donors by Mn acceptors, we may conclude that the concentration of $As_{Ga}$ is equal to 0.15% at the 100% compensation point. It is half of the $X_{Mn}$ since $As_{Ga}$ is a double donor.

In order to verify the estimations of Mn interstitials concentrations in GaMnAs with different concentrations of As antisites, we used a procedure which is recognized to remove $Mn_I$ defects (but not affecting $As_{Ga}$), namely the low temperature post-growth annealing[20]. Fig. 4 shows the annealing effects on both LT GaAs buffers and GaMnAs layers of samples 1, 2 and 3. As seen in Fig. 4 a, b, c the effect of annealing on the lattice constant of LT GaAs layers is negligible. The LT GaAs lattice parameter does not change almost at all even after the highest temperature annealing. That means that the defects present in LT GaAs are not affected by annealing to 280 °C and below. This is consistent with the literature reports[35] indicating that LT GaAs changes its defect structure upon annealing to much higher temperatures - above 400 °C. In the case of GaMnAs layers, the influence of annealing on $a_{GaMnAs}$ is significant. The most interesting effect is the decrease of annealing induced changes in $a_{GaMnAs}$ with increasing value of the excess As flux used during GaMnAs LT MBE growth. For sample 3, grown at the highest $r_{As/Ga}$ a slight *increase* of $a_{GaMnAs}$ after annealing was observed. This is in contrast to what is observed for the samples 1 and 2, and to the observations reported by other groups[36,37], which all show decreased lattice constant upon annealing. The annealing induced decrease of the GaMnAs lattice constant can be interpreted as an effect of removing Mn from interstitial sites. This was suggested theoretically[32] and recently shown experimentally[37]. Our results showing the disappearance of this effect for samples with high density of As antisites suggest that increasing density of As antisites in GaMnAs is accompanied by the decreasing density of Mn interstitials. Simple analyze of the influence of Mn interstitials and As antisites on GaMnAs lattice parameter based on the results of theoretical model by Masek et. al.[32] confirms this conclusion. This is also in agreement with the recent theoretical work of M. Mahadevan and A. Zunger[38], who found that the MBE growth of GaMnAs at high excess As conditions inhibits formation of $Mn_I$ defects. However the As rich growth conditions promote the formation of As antisites which are efficient compensating centers that can not be removed by the post-growth annealing, in contrast to the Mn interstitials.

In summary, we have shown that the lattice constant of GaMnAs depends on the concentration of both: As antisite and Mn interstitial defects. At the very low Mn concentrations, the GaMnAs lattice constant slightly decreases with increasing Mn content, up to the Mn content at which As antisite donors are fully compensated by the Mn acceptors. Further increase of Mn concentration leads to the increase of $a_{GaMnAs}$, due to the manganese at gallium sites and increased density of manganese at interstitial sites. The lattice constant measurements of GaMnAs with different concentrations of As antisites as well as the results of low temperature annealing experiments indicate that there is a balance between Mn interstitials and As antisite defects during the low temperature MBE growth process of GaMnAs. This is leading to the reduced density of one type of defect upon increased density of another defect in as grown GaMnAs films.

The authors would like to thank Dr Janusz Kanski from Chalmers University of Technology, Göteborg, Sweden for valuable discussions. This work was supported in part by the European Commission program ICA1-CT-2000-70018 (Centre of Excellence CELDIS).

**Table 1**. Parameters of as grown, non annealed $Ga_{0.96}Mn_{0.04}As$ samples and LT GaAs buffer layers.

| Sample No | LT GaAs thickness [μm] | GaMnAs thickness [μm] | LT GaAs $a_{strained}$ $a_{relaxed}$ [Å] | GaMnAs $a_{strained}$ $a_{relaxed}$ [Å] | [$As_{Ga}$] | [$Mn_I$] |
|---|---|---|---|---|---|---|
| 1 | 0.80 | 0.50 | 5.6564 5.6549 | 5.67530 5.66408 | 0.20% | 0.82% |
| 2 | 0.20 | 0.30 | 5.65688 5.6551 | 5.67316 5.66305 | 0.23% | 0.70% |
| 3 | 0.20 | 0.40 | 5.6587 5.6560 | 5.67358 5.66325 | 0.35% | 0.64% |

**Figure captions**

**Fig. 1.** (004) X-ray Bragg reflections for two $Ga_{0.999}Mn_{0.001}As$ layers grown at: (a) - high excess As flux; (b) - low excess As flux. Peak on the right side – reflection from the GaAs(001) substrate, peak on the left side – reflection from the GaMnAs layer

**Fig. 2.** Perpendicular (strained) lattice constant of $Ga_{1-x}Mn_xAs$ with Mn content x from 0.1% to 3.5%. At x=0.3% $a_{strained}$ reaches the lowest value.

**Fig. 3.** (006) X-ray Bragg reflections for three GaMnAs samples grown on LT GaAs buffers at different $As_2$/Ga flux ratios of: 2 – sample #1, solid line; 5 – sample #2, dashed line; 9 – sample #3, dotted line

**Fig. 4.** (006) X-ray Bragg reflections for: samples 1, 2 and 3 (Figs. 4a, 4b, 4c) before and after post-growth annealing. Solid, dashed, short dashed and dotted lines correspond to the samples: non annealed, annealed at 240 °C, annealed at 260 °C, annealed at 280 °C, respectively.

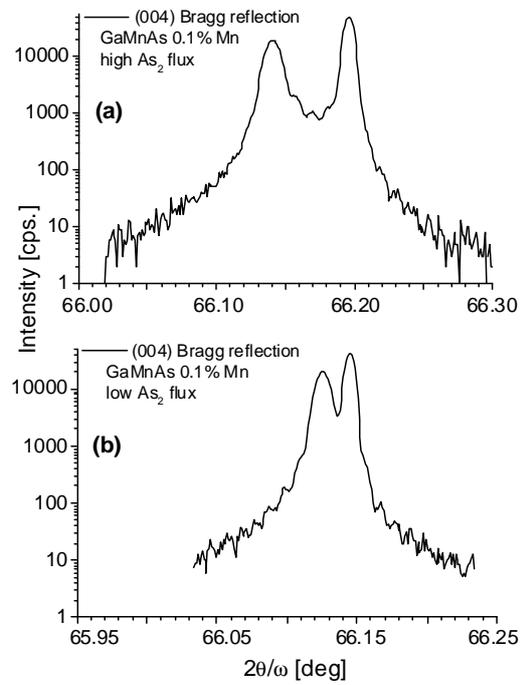

*Sadowski and Domagala,* Fig. 1

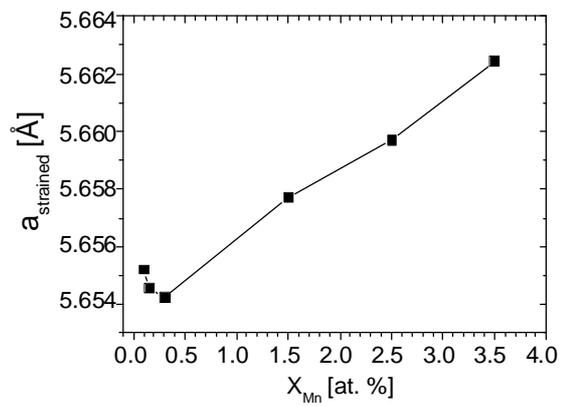

*Sadowski and Domagala,* Fig. 2

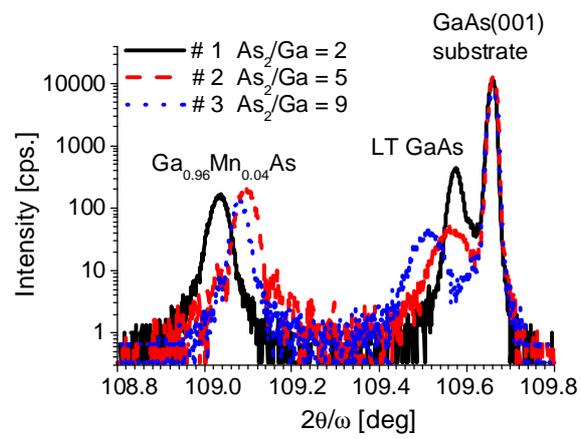

*Sadowski and Domagala,* Fig. 3

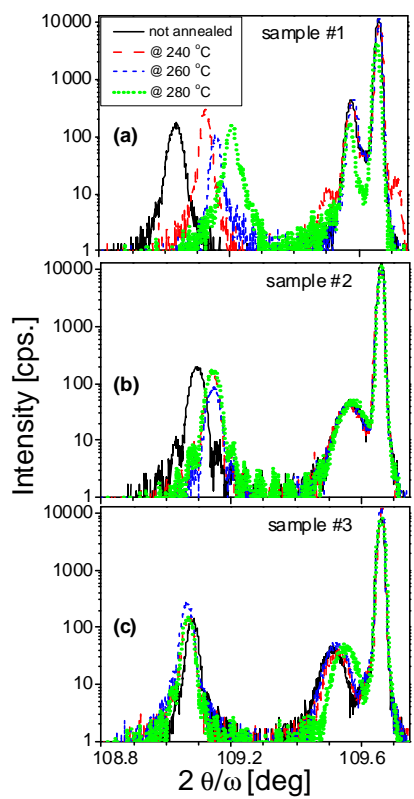

*Sadowski and Domagala,* Fig. 4